\documentclass[conference]{IEEEtran}
\usepackage{graphicx}
\usepackage{array}
\usepackage{layout}
\usepackage{bm}
\usepackage{algorithmic}
\usepackage{amssymb, amsmath}
\usepackage[numbers, square, comma, sort&compress]{natbib}
\usepackage[ruled,lined]{algorithm2e}
\usepackage{epstopdf}
\usepackage{textgreek}
\usepackage{subcaption}
\usepackage{siunitx}
\usepackage{url}
\usepackage{xcolor}
\usepackage[english]{babel}
\usepackage{multicol}
\usepackage{color}
\usepackage{acronym}
\usepackage{mathtools}
\usepackage[verbose]{wrapfig}
\usepackage[normalem]{ulem}
\usepackage [autostyle, english = american]{csquotes}
\MakeOuterQuote{"}
\usepackage{caption} 
   \usepackage[belowskip=2pt, aboveskip=-13pt] {caption}
\DeclarePairedDelimiter{\ceil}{\lceil}{\rceil}

\newcommand{\eq}[1]{(\ref{#1})}

\acrodef{MEC}[MEC]{Multi-access Edge Computing}
\acrodef{BS}[BS]{Base Station}
\acrodef{ML}[ML]{Machine Learning}
\acrodef{EB}[EB]{Energy Buffer}
\acrodef{EH}[EH]{Energy Harvesting}
\acrodef{QoS}[QoS]{Quality of Service}
\acrodef{MN}[MN]{Mobile Network}
\acrodef{API}[API]{Application Programmable Interface}
\acrodef{ES}[ES]{Energy Saving}
\acrodef{VM}[VM]{Virtual Machine}
\acrodef{EM}[EM]{Energy Manager}
\acrodef{LLC}[LLC]{Limited Lookahead Control}
\acrodef{LSTM}[LSTM]{Long Short-Term Memory}
\acrodef{CPU}[CPU]{Central Processing Unit}
\acrodef{LS}[LS]{Location Service}
\acrodef{VNF}[VNF]{Virtualized Network Function}
\acrodef{NFV}[NFV]{Network Function Virtualization}

\IEEEoverridecommandlockouts\IEEEpubid{\makebox[\columnwidth]{978-1-7281-2294-6/19/\$31.00~\copyright 2019 IEEE \hfill} \hspace{\columnsep}\makebox[\columnwidth]{ }} 

\begin{document}

\title{\ Adaptive Resource Management for a Virtualized Computing Platform within Edge Computing}

\author{\IEEEauthorblockN {Thembelihle Dlamini\IEEEauthorrefmark{1}\IEEEauthorrefmark{2}, \'Angel Fern\'andez Gamb\'in\IEEEauthorrefmark{1}}
	\IEEEauthorblockA {\IEEEauthorrefmark{1}Department of Information Engineering, University of Padova, Padova, Italy}
	\IEEEauthorblockA {\IEEEauthorrefmark{2}Athonet, Bolzano Vicentino, Vicenza, Italy}
	\{dlamini, afgambin\}@dei.unipd.it
	 \vspace{-0.4cm}
}
\maketitle

\begin{abstract}

In virtualized computing platforms, energy consumption is related to the \mbox{\it computing-plus-communication} processes. However, most of the proposed energy consumption models and energy saving solutions found in literature consider only the active \acp{VM}, thus the overall operational energy expenditure is usually related to solely the computation process. 
To address this shortcoming, in this paper we consider a \mbox{computing-plus-communication} energy model, within the {\it\ac{MEC}} paradigm, and then put forward a combination of a traffic engineering- and \ac{MEC} Location \mbox{Service-based} online server management algorithm with {\it \ac{EH}} capabilities, called Automated Resource Controller for \mbox{Energy-aware} Server (ARCES), for autoscaling and reconfiguring the \mbox{computing-plus-communication} resources. The main goal is to minimize the overall energy consumption, under hard \mbox{per-task} delay constraints (i.e., \ac{QoS}). 
ARCES jointly performs (i) a \mbox{short-term} server demand and harvested solar energy forecasting, (ii) \ac{VM} \mbox{soft-scaling}, workload and processing rate allocation and lastly, (iii) switching on/off of transmission drivers (i.e., {\it fast} tunable lasers) coupled with the \mbox{location-aware} traffic scheduling.
Our numerical results reveal that ARCES achieves on average energy savings of $69$\%, and an energy consumption ranging from $31$\%-$45$\%and from $21$\%-$25$\% at different values of \mbox{per-VM} reconfiguration cost, with respect to the case where no energy management is applied.

\end{abstract}


\IEEEpeerreviewmaketitle

\section{Introduction}

The data growth generated by pervasive mobile devices and the Internet of Things, couple with the demand for \mbox{ultra-low} latency, requires high computation resources which are not available at the \mbox{end-user} device. Undoubtedly, offloading to a powerful computational \mbox{resource-enriched} server located closer to mobile users is an ideal solution. Thus, \mbox{Multi-access} Edge Computing (MEC~\cite{etsimec_access}) has recently emerged to enable \mbox{low-latency} and \mbox{location-aware} data processing at the {\it network edge}, i.e., in close proximity to mobile devices, sensors, actuators and connected things. Despite the potential presented by MEC, the computing resources (i.e., \acp{VM}) plus communication within the MEC server (node) raises concerns related to energy consumption, in the effort to build greener networks and reduce carbon emissions into the atmosphere. Nonetheless, with the integration of \ac{EH} into the edge (or computing) system~\cite{energymanagershow}, resulting into an \mbox{EH-powered} MEC (\mbox{EH-MEC}) system, the carbon footprint and the dependence on the power grid can be minimized.

The energy drained in the computing platform due to the \mbox{computing-plus-communication} processes is associated with (i) the running \acp{VM}~\cite{virttech}\cite{eempirical} and (ii) the communication within the server's virtual network~\cite{vm_book} ({\it see} Fig. 3.5 in this reference). It is observed in the literature that most of the existing \ac{ES} studies have involved autoscaling~\cite{xu2016online}\cite{online_pimrc}\cite{shojafar2015energy}\cite{vm_char}\cite{guenter2011managing} (scaling up/down the number of computing nodes/servers or \acp{VM}), \ac{VM} migration~\cite{Beloglazov} (movement of a \ac{VM} from one host to another) and soft resource scaling~\cite{soft_virtual} (shortening the access time to physical resources), all hereby referred to as \ac{VM} \mbox{\it soft-scaling}, i.e., the reduction of computing resources per time instance. Hence, the proposed energy models (see~\cite{power_metering} for a summary of the proposed models) and the overall operational expenditure of the computing node is usually related to the computation process (i.e., the running VMs), overlooking the communication processes within the server.

Regarding energy consumption related to communication within a computing node, it is shown in~\cite{link_drivers}\cite{laser_tuning} that having the least number of data transmission drivers ({\it fast} tunable lasers) can yield significant amount of ESs. It is worth observing that both works,~\cite{link_drivers}\cite{laser_tuning}, are not along the direction of MEC, but they propose the tuning of the transmission drivers as {\it one} of the ES strategies within the \ac{MN} infrastructure. Thus, ESs within the MEC server can be {\it jointly} achieved by launching an optimal number of VMs for computing, and transmission drivers {\it coupled} with the \mbox{location-aware} traffic routing for \mbox{real-time} data transfer. 

\subsection{Motivation} 

The virtualized MEC node is equipped with higher computational and storage resources compared to the \mbox{end-users} devices, in order to handle the computation workload being generated at the network edge. However, while MEC tries to meet the computational demand and the guarantee of \mbox{low-latency}, the issue of energy consumption is still a challenge within the virtualized computing node~\cite{virttech}\cite{eempirical}. To address this challenge: (i) it is expected that the \ac{NFV} framework can exploit the benefits of virtualization technologies to significantly reduce the energy consumption in MN infrastructure; (ii) the current trends in battery and solar module costs show an expected reduction. This two points motivate the integration of \ac{MEC} and {EH} systems towards green computing~\cite{energymanagershow}\cite{nicola}.

\subsection{Related work}

For several years, great effort has been devoted to study energy savings in computing environments with the aim of minimizing the energy consumption. Procedures for the dynamic on/off switching of servers have been proposed as a way of minimizing energy consumption in computing platforms. In~\cite{xu2016online}, computing resources are provisioned depending on the expected server workloads via a reinforcement \mbox{learning-based} resource management algorithm, which learns the optimal policy for dynamic workload offloading and servers autoscaling. 
Our previous works in~\cite{online_pimrc} and~\cite{edge_controller}, focus on  the provision of computing resources (VMs) based on a \ac{LLC} policy and the network impact (the use of traffic load as a performance metric~\cite{E_oh}), after forecasting the future workloads and harvested energy. A single \ac{BS} optimization case is considered for an \mbox{off-grid} site in~\cite{online_pimrc}, and a multiple \ac{BS} optimization case, each \ac{BS} site powered by hybrid energy sources, is studied in~\cite{edge_controller} where the edge management procedures are enabled by an edge controller. This work differs from our previous works as the MEC server is placed in proximity to a \ac{BS} cluster, and not one \mbox{co-located} for each \ac{BS}. Moreover, here we focus on the integration of \mbox{communication-related} energy consumption by considering the tuning of the transmission drivers, which is a novel concept within the \ac{MEC} paradigm.
In~\cite{Beloglazov}, \ac{CPU} utilization thresholds are used to identify \mbox{over-utilized} servers. \acp{VM} are migrated to servers that will accept them without incurring in high energy costs. Subsequently, the idle servers are \mbox{turned-off}.

Energy management is also of interest in data centers using virtualization technologies. Along the same lines of \ac{VM} \mbox{soft-scaling}, in~\cite{shojafar2015energy} a traffic  \mbox{engineering-based} adaptive approach is presented with the aim of minimizing energy consumption induced by computing, communication and reconfiguration costs of virtualized clouds. An iterative method is used to obtain \acp{ES} within a server that transmits wirelessly to clients. Then, in~\cite{vm_char} the \mbox{computing-plus-communication} is also considered towards a goal of saving energy through an adaptive transmission rate for a Fog node.
An automated server provisioning algorithm that aims to meet workload demand while minimizing energy consumption in data centers is presented in~\cite{guenter2011managing}. Here, \mbox{energy-aware} server provisioning  is performed by taking into account \mbox{trade-offs} between cost, performance, and reliability.
Lastly, a {\it soft resource scaling} mechanism is proposed in~\cite{soft_virtual} where the scheduler shortens the maximum resource usage time for each \ac{VM}, i.e., the time slice allocated for using the underlying physical resources, in order to compensate for the low energy savings achieved with Dynamic Voltage and Frequency Scaling (DVFS).

\subsection{Objective and Contributions}
\label{sub:obj_contr}

The main contributions of this work are as follows: 

\begin{itemize}
	\item we consider the aforementioned scenario, where \ac{MEC} and \ac{EH} are combined into a single system located close to a \ac{BS} cluster, towards energy \mbox{self-sustainability} in MNs. The \mbox{EH-MEC} system is equipped with solar panels for \ac{EH} and an \ac{EB} for energy storage. 
	
	\item We consider a \mbox{computing-plus-communication} energy model within the {\it\ac{MEC}} paradigm, formulating a constrained optimization problem. Due to the \mbox{non-linear} behavior of the \mbox{rate-vs-power} relationship, the optimization problem is \mbox{non-convex}. To solve it, we convexify the function by using Geometric Programming (GP~\cite{geo_prog}) and then employing the CVXOPT toolbox\footnote{M. Andersen and J. Dahl. CVXOPT: Python Software for Convex Programming, 2019. [Online]. Available: https://cvxopt.org/} and approximations.
	
	\item  We forecast the \mbox{short-term} future server workload and harvested energy, by using a \ac{LSTM} neural network~\cite{lstmlearn}, to enable foresighted optimization.
	
	\item Lastly, we develop an online \mbox{controller-based} algorithm {\it called} Automated Resource Controller for \mbox{Energy-aware} Server (ARCES) for the MEC server management based on \ac{LLC} theory~\cite{hayes} and energy management procedures. The main goal is to minimize the overall energy consumption, under hard \mbox{per-task} delay constraints (i.e., \ac{QoS}), through the {\it joint} consideration of \ac{VM} \mbox{soft-scaling} and the tuning of transmission drivers coupled with the \mbox{location-aware} traffic routing. To the best of our knowledge, this is a novel concept within the \ac{MEC} paradigm. ARCES considers future server workloads, onsite green stored energy, and target \ac{BS} (based on the \ac{LS}~\cite{etsimec_ls}), and then enable \ac{ES} procedures.
\end{itemize}

The proposed optimization strategy is able to reduce the energy consumption under the guidance of the online resource controller and the energy management procedures.

The rest of the paper is organized as follows. The system model is presented in Section~\ref{sec:sys}. In Section~\ref{sec:prob}, we detail the optimization problem and the proposed \mbox{\ac{LLC}-based} online algorithm. Simulation results are discussed in Section~\ref{sec:per_results}. Lastly, we conclude our work in Section~\ref{sec:concl}.

\begin{figure} [t]
	\centering
	\includegraphics[width = 0.48\textwidth]{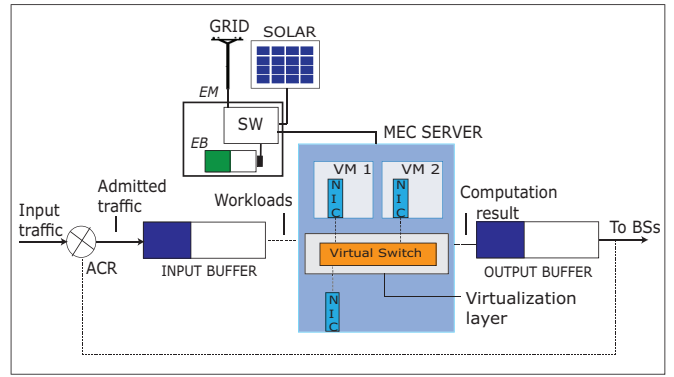}
	\caption{Virtualized computing system powered by hybrid energy sources: \mbox{on-grid} power and green energy. The electromechanical switch (SW) selects the appropriate source of energy.}
	\label{fig:edge_sys}
\end{figure} 

\section{System Model}
\label{sec:sys}

As a major deployment of MEC~\cite{etsimec_access}, the considered network scenario is illustrated in Fig.~\ref{fig:edge_sys}. It consists of a \mbox{cache-enabled}, TCP/IP \mbox{offload-enabled} ({\it partial} computation at the network adapter), virtualized MEC server hosting $M$ VMs and it is assumed to be deployed at an aggregation point~\cite{etsimec_access}\cite{energymanagershow}, i.e., a point in proximity to a group of \acp{BS} interconnected to the MEC server for computation offloading. The MEC node is assumed to be equipped with higher computational and storage resources compared to the \mbox{end-user} device.
The server clients are assumed to be mobile users moving in groups and they are represented by the Reference Point Group Mobility Model (RPGM~\cite{mobility_dir}). Their current locations are known through the \ac{LS} \ac{API}~\cite{etsimec_ls}, in the MEC platform, which is a service that supports UE's location retrieval mechanism, and then passing the information to the authorized applications within the server.
The computing site is empowered with \ac{EH} capabilities through a solar panel and an \ac{EB} that enables energy storage. Energy supply from the power grid is also available for backup.
The \ac{EM} is an entity responsible for selecting the appropriate energy source and for monitoring the energy level of the \ac{EB}. The virtualized Access Control Router (ACR) of Fig.~\ref{fig:edge_sys} acts as an access gateway, responsible for routing, and it is locally hosted as an application. Moreover, we consider a \mbox{discrete-time} model, whereby time is discretized as \mbox{$t = 1,2,\dots$}, and each time slot $t$ has a fixed duration $\tau$.  

\subsection{Server Workload and Energy Consumption}
\label{sub:serverload}

\begin{figure}[t]
	\centering
	\resizebox{\columnwidth}{!}{\input{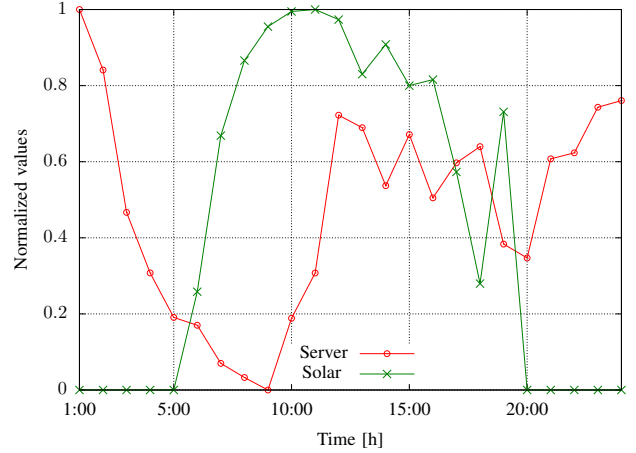}}	
	\caption{Example traces for server workloads from~\cite{workload_traces} and harvested solar energy from~\cite{amerinia}.}
	\label{fig:profiles}
\end{figure}

For many MN services, the workload demand exhibits a diurnal behavior, thus it suffices to forecast the \mbox{short-term} server workload (using historical datasets~\cite{energymanagershow}\cite{guenter2011managing}) and then enable dynamic resource management within the server. In this work, anonymized real server workload traces obtained from~\cite{workload_traces} are used to emulate server workloads due to the difficulties in obtaining relevant open source datasets containing computing requests. 
A trace file consist of the file size, session duration, total number of packets and average transmission rate, over one day. In our numerical results, we use the total number of packets, denoted by $L_{\rm in}(t)$ ([bits]), to represent the buffered (or admitted) computation workload at the input buffer at time slot $t$ ({\it see} red curve in Fig.~\ref{fig:profiles}). In addition, we assume that the \mbox{upper-bounded} input/output (I/O) queue's of Fig.~\ref{fig:edge_sys} are \mbox{loss-free} and they implement the \mbox{First-In First-Out} (FIFO) service discipline, thus $L_{\rm in}(t) = L_{\rm out}(t)$, where $L_{\rm out}(t)$ is the amount of the aggregate computation result stored at the output buffer. 

The total energy consumption ([$\SI{} {\joule}$]) for the virtualized computing platform is formulated as follows, inspired by~\cite{online_pimrc}\cite{shojafar2015energy} and the virtualization knowledge from~\cite{vm_book}:
\begin{equation}
	\mbox{$\theta_{\rm MEC}(t) = \theta_{\rm COMP}(t) + \theta_{\rm COMM}(t)$} \, ,         
	\label{eq:mecconsupt}
\end{equation}
\noindent where $\theta_{\rm COMP}(t)$ is the energy drained due to computation and $\theta_{\rm COMM}(t)$ due to \mbox{intra-communications} processes in the MEC server at time slot $t$.\\ 

\subsubsection*{\bf Computing energy} $\theta_{\rm COMP}(t) =  \theta_{\rm CPU}(t) + \theta_{\rm SC}(t) + \theta_{\rm TOE}(t)$, where $\theta_{\rm CPU}(t)$ is the energy drained due to the running \acp{VM}, w.r.t \ac{CPU} utilization, and $\theta_{\rm SC}(t)$ is the energy drained due to VM switching the processing rates $f_m(t) \in [0, f_{\rm max}]$. $f_{\rm max}$ [(bit/s)] is the maximum processing rate for \ac{VM} $m$. $\theta_{\rm TOE}(t)$ is the energy induced by the TCP/IP offload on the network interface card (NIC, e.g., TCP/IP checksum offload).
In practice, the VMs are instantiated on top of the CPU cores and each VM processes the currently allotted task by managing its own local virtualized computing resources, thus we model the processing rates to be between $f_0 = 0$ (represents zero speed of the \ac{VM}, e.g., deep sleep or shutdown) and $f_{\rm max}$.
Here, we assume that \mbox{real-time} processing of computation workloads is performed in parallel over the VMs interconnected by a \mbox{power-limited} and \mbox{rate-adaptive} switched Virtual Local Area Network (VLAN). 

Considering that $\theta_{\rm CPU}(t)$ is related to the number of \acp{VM} running in time slot $t$, named $M(t) \leq M$, and on the \ac{CPU} frequency that is allotted to each VM, $\theta_{\rm CPU}(t)$ is obtained using the linear relationship between the \ac{CPU} utilization contributed by VM $m$, and the energy drained is, inspired by~\cite{online_pimrc}: 
\begin{equation}
     \mbox{$\theta_{\rm CPU}(t) = \sum_{m=1}^{M(t)}\theta_{{\rm idle}, m}(t) +  \theta_{{\rm dyn},m}(t)$},
    \label{eq:vm_cpu}
\end{equation}
where $\theta_{{\rm idle}, m}(t)$  represents the {\it static} energy drained by VM $m$ in the idle state, and the quantity $\theta_{{\rm dyn},m}(t) = \alpha_{m}(t) (\theta_{{\max}, m}(t) - \theta_{{\rm idle}, m}(t))$ represents the {\it dynamic} energy component of VM $m$, where \mbox{$\alpha_{m}(t) = (f_{m}(t)/f_{\max})^{2}$}~\cite{hayes} is a load dependent factor and $\theta_{{\max}, m}(t)$ is the {\it maximum} energy that VM $m$ can drain.  

Next, we remark that the VM switching cost $\theta_{\rm SC}(t)$ depends on the frequency reconfiguration, i.e., the transition from $f_1(t)$ (current processing rate for VM $m$) to $f_2(t)$ (the next processing rate), as an example. In short, the energy cost depends on the absolute processing rate gap, $|f_2(t) - f_1(t)|$. Thus, $\theta_{\rm SC}(t)$ is defined as~\cite{shojafar2015energy}:
\begin{equation}
	\mbox{$\theta_{\rm SC}(t) = \sum_{m=1}^{M(t)} \kappa_e(f_2(t) - f_1(t))^2$},
	\label{eq:vm_sc}
\end{equation}
where $\kappa_e$ is the the \mbox{per-VM} reconfiguration cost caused by a \mbox{unit-size} frequency switching. Typically, $\kappa_e$ is limited to a few hundreds of $\SI{} {\milli\joule}$ per $(\SI{} {\mega\hertz})^{2}$.

At this regard, we put forward the following: at the beginning of time slot $t$, the online resource controller adaptively allocates the available virtual resources and thus determines the VMs demanded, $M(t)$, the workload allotted to \ac{VM} $m$, denoted by $\lambda_{m}(t)$, and $f_m(t)$ for VM $m$ that will yield the desired or expected processing time, \mbox{$\chi_{m}(t) = \lambda_{m}(t)/f_{m}(t)$}. Note that \mbox{$L_{\rm in}(t) = \sum_{m = 1}^{M(t)} \lambda_m(t)$}. Moreover, in practical application scenarios, the maximum \mbox{per-VM} computation load to be computed is generally limited up to an assigned value, named $\lambda_{\rm max}$. 
Lastly, the VM provisioning and workload allocation is discussed in Section~\ref{server_manager}, and $f_m(t) \stackrel{\Delta}{=} \lambda_m(t) / \Delta$.

Along the same lines of computation, advancement in TCP/IP Offload Engine (TOE) technology enables {\it partial} computation in the server's NIC~\cite{sohan2010characterizing}, i.e., some TCP/IP processing (e.g., checksum computation) is offloaded to a specialized hardware on the network adapter, relieving the host CPU from the overhead of processing TCP/IP. 
Thus, $\theta_{\rm TOE}(t)$ is obtained by using the fact that it is data volume dependent and then determined using the workload volume received. Due to the lack of an existing TOE energy model, we rely on the performance measure for the Broadcom (Fibre) $10$ Gbps NIC~\cite{sohan2010characterizing}, which is considered here as an example of a TCP/IP \mbox{offload-capable} device. Then, $\theta_{\rm TOE}(t)$ is obtained as:
\begin{equation}
    \mbox{$\theta_{\rm TOE}(t) = \zeta(t)\,\theta_{\rm idle}^{\rm TOE}(t) \, + \theta_{\rm max}^{\rm TOE}(t)$},
  \label{eq:vm_toa}
\end{equation} 
where $\theta_{\rm idle}^{\rm TOE}(t) > 0$ is the energy drained by the TOE when powered, with all links connected without any data transfer. This motivates the idea of tuning even the NIC so that the energy drained is always zero when there is no data transfer. For this, we have $\zeta(t) = (0,1)$ as the NIC switching status indicator ($1$ for active state and $0$ for idle state). $\theta_{\rm max}^{\rm TOE}(t) = \frac{L_{\rm in}(t)}{\eta}$ is the maximum energy drained by the TOE. $\eta$ is a fixed value measured in [Gbit/$\SI{} {\joule}$]. \\

\subsubsection*{\bf Communication energy} $\theta_{\rm COMM}(t)= \theta_{\rm VLAN}(t) + \theta_{\rm WCOM}(t)$, where $\theta_{\rm VLAN}(t)$ is the energy drained due to the communication links (to-and-from each VM), and $\theta_{\rm WCOM}(t)$ is the energy drained due to the number of transmission (optical) drivers used for the data transfer to target BS(s). 

The communication energy within the VLAN is obtained by using the \mbox{Shannon-Hartley} exponential analysis. Here, we assume that each VM $m$ communicates with the resource controller through a dedicated reliable link, that operates at the transmission rate of $r_m(t)$ [(bit/s)]. Thus, the energy needed for sustaining the \mbox{two-way} $m^{\rm th}$ link is defined as, inspired by~\cite{nicola}:
\begin{equation}
	\mbox{$\theta_{\rm VLAN}(t) = 2\,\sum_{m=1}^{M(t)} P_m(r_m(t))(\lambda_m(t)/ r_m(t))$},
	\label{eq:vm_vlan}
\end{equation}
where \mbox{$P_m(r_m(t)) = \Gamma_m (2^{r_m(t)/W_m} - 1)$} is the power drained by the $m^{\rm th}$ communication link and \mbox{$\Gamma_m = \frac{W_m \times N_0^{(m)}}{g_m}$}. $N_0^{(m)}(\SI{} {\watt /\hertz})$ is the noise spectral power density, $W_m$ is the bandwidth, and $g_m$ is the (\mbox{non-negative}) gain of the $m^{\rm th}$ link. In practical application scenarios, the maximum \mbox{per-slot} communication rate within the \mbox{intra-VLAN} is generally limited up to an assigned value $r_{\rm max}$. Thus, the following hard constraint must hold: $ \sum_{m=1}^{M(t)} r_m(t) \leq r_{\rm max}$.

Before proceeding, we consider the \mbox{\it two-way} \mbox{per-task} execution delay ([s]). We have the  $m = \{1,\dots, M(t)\}$ link connection delays, each denoted by $\Omega_m(t) = \lambda_m(t)/r_m(t)$, and $\chi_m(t) \leq \Delta$, where $\Delta$ is the maximum \mbox{per-slot} and \mbox{per-VM} processing time ([s]). At this regard, we note that $\Delta$ is also the server's response time, i.e., the maximum time allowed for processing the total computation load and it is fixed in advance regardless of the task size allocated to VM $m$. Since parallel \mbox{real-time} processing is assumed in this work, the overall communication equates to $2\,\Omega_{m}(t) + \Delta$. Therefore, the hard \mbox{per-task} delay constraint on the computation time is: $\max \{2\,\Omega_{m}(t)\} + \Delta \leq \tau_{\rm max}$, where $\tau_{\rm max}$ is the maximum tolerable delay, which is fixed in advance.

Finally, $\theta_{\rm WCOM}(t)$ depends on the number of laser (optical) drivers, named $Y(t) \leq Y$ ($Y$ is the total number of them), that are required for transferring $ \ell_y(t) \in L_{\rm out}(t)$ in time slot $t$ ($\ell_y(t)$ is the downlink traffic volume ([bits] of the driver at slot $t$). $L_{\rm out}(t)$ is accumulated over a fixed period of time to form a {\it batch} at the output buffer. 
At this regard, we note that a large number of drivers yield large transmission speed while at the same time resulting into high energy consumption~\cite{laser_tuning}. Therefore, the energy consumption can be minimized by launching an optimal number of drivers for the data transfer. Moreover, for every {\it mobile client} who offloaded their task into the MEC server associated with the radio nodes, i.e. BSs, its location and the computation result is known through the UE subscription procedure (i.e., through the \mbox{LS}), thus enabling the \mbox{location-aware} traffic routing and obtaining $Y(t)$.

The energy drained during the data transmission process consists of the following: a constant energy for utilizing each fast tunable driver denoted by $O_{{\rm opt},y}(t)$ ([$\SI{} {\joule}$/s]), the target transmission rate $r_{0}$ [bits/s] and $ L_{\rm out}(t)$. Thus, the energy is, inspired by~\cite{link_drivers}\cite{chen2018computation}: 
\begin{equation}
	\mbox{$\theta_{\rm WCOM}(t) = \sum_{y=1}^{Y(t)}\frac{O_{{\rm opt},y}(t)\,l_y(t)}{r_0}$},
	\label{eq:vm_wcom}
\end{equation}
\noindent where the parameter $Y(t)$ is obtained using the total number of target \acp{BS} as $Y(t) = \ceil[\big]{\frac{1}{\alpha}\cdot(\frac{\omega (t) + 1}{\omega(t)})^2}$ (see~\cite{link_drivers}), where $\omega(t) = \sqrt{\frac{\Upsilon}{\sigma N_{\rm BS}(t)}}$. $\alpha \in (0,1]$ is a controllable factor that determines the delay constraint of optical networks, $\sigma$ ([$\SI{} {\milli\second}$]) is the reconfiguration cost for tuning the transceivers, $N_{\rm BS}(t)$ is an integer value representing the total number of target \acp{BS} at time slot $t$, and $\Upsilon$ is the number of time slots at which the computed workload is accumulated at the output buffer. $\alpha, \sigma,$ and $\Upsilon$ are fixed values. $ L_{\rm out}(t)$ is equally distributed over the $Y(t)$ drivers.

\subsection{Energy Patterns and Storage}
\label{sub:eebuffer}

The energy buffer of Fig.~\ref{fig:edge_sys} is characterized by its maximum energy storage capacity $B_{\rm max}$, and power charging/discharging and leaking losses are not assumed. At each time slot $t$, the \ac{EM} provides the energy level report to the MEC server, through the pull mode procedure (e.g., File Transfer Protocol~\cite{filetransfer}), thus the \ac{EB} level $B(t)$ is known, enabling the provision of the required computation and communication resources, i.e., the \acp{VM} and laser drivers.

In this work, the amount of harvested energy $H(t)$ in time slot $t$ is obtained from \mbox{open-source} solar traces within a solar panel farm located in Armenia~\cite{amerinia} ({\it see} green curve in Fig.~\ref{fig:profiles}), where the dataset time scale matches our time slot duration ($\SI{1} {\minute}$). The dataset is the result of daily environmental records for a place assumed to be free from surrounding obstructions (e.g., buildings, shades). In our numerical results, $H(t)$ is obtained by picking one day data from the dataset and then scaling the solar energy to fit the \ac{EB} capacity $B_{\rm max}$ of $\SI{490} {\kilo\joule}$. Thus, the available \ac{EB} level $B(t + 1)$ at the beginning of time slot $t+1$ is calculated as follows:  
\begin{equation}
B(t + 1) = B(t) + H(t) - \theta_{{\rm MEC}}(t) + E(t),
\label{eq:offgrid}
\end{equation}
\noindent where $B(t)$ is the energy level in the battery at the beginning of time slot $t$, $\theta_{\rm MEC}(t)$ is the energy consumption of the computing platform over time slot $t$, see Eq.~(\ref{eq:mecconsupt}), and $E(t) \geq 0$ is the amount of energy purchased from the power grid. We remark that $B(t)$ is updated at the beginning of time slot $t$ whereas $H(t)$ and $\theta_{\rm MEC}(t)$ are only known at the end of it.

For decision making by the resource controller, the received \ac{EB} level reports are compared with the following thresholds: \mbox{$B_{\rm low}$} and $B_{\rm up}$, respectively termed the lower and the upper energy threshold with \mbox{$0 < B_{\rm low} < B_{\rm up} < B_{\rm max}$}. \mbox{$B_{\rm up}$} corresponds to the desired energy buffer level and \mbox{$B_{\rm low}$} is the lowest EB level that the MEC server should ever reach. The suitable energy source at each time slot $t$ is selected based on the forecast expectations, i.e., the expected harvested energy $\hat{H}(t)$. If $\hat{H}(t)$ is enough to reach \mbox{$B_{\rm up}$}, no energy purchase is needed. Otherwise, the remaining amount up to \mbox{$B_{\rm up}$}, i.e., \mbox{$E(t) = B_{\rm up} - B(t)$} is bought from the electrical grid. Our optimization framework in Section~\ref{sub:opt_prob} makes sure that $B(t)$ never falls below $B_{\rm low}$ and guarantees that \mbox{$B_{\rm up}$} is reached at every time slot.

\section{Problem Formulation}
\label{sec:prob}

In this section, we formulate an optimization problem to obtain reduced energy consumption through \mbox{short-term} server workload and harvested solar energy forecasting along with server management procedures. The optimization problem is defined in Section~\ref{sub:opt_prob}, and the online server management procedures are presented in Section~\ref{sub:server_manager}. 

\subsection{Optimization Problem}
\label{sub:opt_prob}

On per time slot basis, the online controller adaptively schedules the communication and computing resources, at the same time receiving the energy level report from the EM. The goal is to minimize the overall resulting \mbox{communication-plus-computing} energy, i.e., the energy consumption related to the MEC server's VMs and transmission drivers. 
To achieve this, for $t=1,\dots, T$, where $T$ is the optimization horizon, we  define the optimization problem as:
\begin{eqnarray}
        \label{eq:objt}
        \textbf{P1} & : & \min_{\mathcal{E}} \mbox{$\sum_{t=1}^T \theta_{\rm MEC}(t)$}\\
        && \hspace{-1.25cm}\mbox{subject to:} \nonumber \\
        {\rm C1} & : & d \leq M(t) \leq M, \nonumber \\
        {\rm C2} & : & B_{\rm low} \leq B(t) \leq B_{\rm max}, \nonumber \\ 
        {\rm C3} & : & 0 \leq f_{m}(t) \leq f_{\rm max}, \nonumber \\
        {\rm C4} & : & 0 \leq \lambda_{m}(t) \leq \lambda_{\rm max}, \nonumber \\
        {\rm C5} & : & \chi_{m}(t) \leq \Delta, \nonumber \\
        {\rm C6} & : & \mbox{$\sum_{m=1}^{M(t)} r_m(t) \leq r_{\rm max}$}, \nonumber\\
        {\rm C7} & : & \max \{2\,\Omega_{m}(t)\} + \Delta \leq \tau_{\rm max}, \nonumber 
\end{eqnarray} 
where $\mathcal{E} \stackrel{\Delta}{=} \{M(t), \{\alpha_m(t)\}, \{P_m(t)\}, \{\lambda_m(t)\}, \zeta(t), Y(t)\}$ is the set of objective variables to be configured at slot $t$ in the MEC server, for the \mbox{computing-plus-communication} processes. Regarding the constraints, C1 forces the required number of \acp{VM}, $M(t)$, to be always greater than or equal to a minimum number \mbox{$d \geq 1$}: the target of this is to be always able to handle mission critical communications. C2 makes sure that the \ac{EB} level is always above or equal to a preset threshold $B_{\rm low}$, to guarantee {\it energy \mbox{self-sustainability}} over time. C3 and C4, bound the maximum processing rate and workloads of each running VM $m$. Constraint C5 represents a \mbox{hard-limit} on the corresponding \mbox{per-slot} and \mbox{per-VM} processing time. Furthermore, C6 bounds the aggregate communication rate sustainable by the VLAN to $r_{\rm max}$ and C7 forces the server to process the offloaded tasks within the set value $\tau_{\rm max}$. 

From P1, we note that $\theta_{\rm MEC}(t)$ consists of a \mbox{non-convex} component, i.e., Eq.~(\ref{eq:vm_vlan}), while the others are convex and \mbox{non-decreasing}. Then, Eq.~(\ref{eq:vm_vlan}) can be convexified into a convex function using GP theory~\cite{geo_prog}, by introducing alternative variables and approximations. In this, we introduce fixed parameters (i.e., $\mu_m, \nu_m$) and approximations. Dropping the index $t$ for convenience, we let \mbox{$r_m = 2\,\lambda_m/(\tau_{\rm max}-\Delta)$}. We obtain $P_m(r_m)$ in terms of $\lambda_m$, by rearranging the \mbox{Shannon-Hartley} expression and substituting the value of $r_m$, as: $\hat{P}_m (r_m)= \frac{((2\,\lambda_m/(\tau_{\rm max} - \Delta)) - \nu_m\,W_m)\ln 2}{\mu_m W_m} + \ln(N_0^{(m)})-\ln g_m$. From the \mbox{Shannon-Hartley} expression, we simply observed the presence of the \mbox{\it log-sum-exp} function as it has been proven to be convex in~\cite{gp_trick} and recall that \mbox{$P_m (r_m) = \exp (\hat{P}_m(r_m))$}.

To solve {\rm P1}, we leverage the use of \ac{LLC}~\cite{hayes}, GP~\cite{geo_prog}, and heuristics, obtaining the feasible system control inputs $\psi(t) = (M(t), \{\alpha_m(t)\}, \{P_m(t)\}, \{\lambda_m(t)\}, \zeta(t), Y(t))$, that yield the best system behavior within $T$. 

\subsection{Resource Controller Design and Server Management}
\label{sub:server_manager}

In this subsection, a server workload and energy harvesting forecasting method, and an online resource management algorithm are proposed to solve the previously stated problem {\rm P1}. In subsection~\ref{predict}, we discuss the \ac{LSTM} neural network used to predict the \mbox{short-term} future server workloads and harvested energy, then in subsection~\ref{server_manager}, we solve {\rm P1} by using \ac{LLC} principles, GP theory, and heuristics, and lastly, in subsection~\ref{alg} we put forward the ARCES algorithm.

\subsubsection{Server workload and energy prediction}
\label{predict}

in order to estimate the system workload over the prediction horizon $T$, we perform time series prediction, i.e., we obtain $T = 3$ estimates of $\hat{\rm L}(t+1)$ and $\hat{H}(t+1)$, by using an \ac{LSTM} network developed in Python using TensorFlow deep learning libraries (Keras, Sequential, Dense, LSTM), with a hidden layer of $4$ \ac{LSTM} neurons, and an output layer that makes a single value prediction. The dataset is split as $67\%$ for training and $33\%$ for testing. As for the performance measure of the model, we use the Root Mean Square Error (RMSE). In this work, prediction steps similar to~\cite{online_pimrc} are adopted (see Table I in this reference), and Fig.~\ref{fig:bs_load} shows the prediction results that will be discussed in Section~\ref{sub:results}. 

\subsubsection{Edge system dynamics}
\label{server_manager}

we denote the system state vector at time $t$ by \mbox{$\textupsilon(t) = (M(t),Y(t),B(t))$}, which contains the number of active VMs, $M(t)$, transmission drivers, $Y(t)$, and the EB level, $B(t)$. The input vector \mbox{$\psi(t) = (M(t), \{\alpha_m(t)\}, \{P_m(t)\}, \{\lambda_m(t)\}, \zeta(t), Y(t))$} drives the MEC server behavior (handles the joint VM \mbox{soft-scaling} and the tuning of transmission drivers) at time $t$. Note that $\{P_m^{*}(t)\}$ is obtained with CVXOPT, and $\{\lambda_m^{*}(t)\}$ is obtained by following {\it remark} $1$. 

The system behavior is described by the \mbox{discrete-time} \mbox{state-space} equation, adopting the \ac{LLC} principles~\cite{hayes}\cite{llcprediction}:
\begin{equation}
\textupsilon(t + 1) = \phi(\textupsilon(t), \psi(t)) \, , 
\end{equation}
\noindent where  $\phi(\cdot)$ is a behavioral model that captures the relationship between $(\textupsilon(t),\psi(t))$, and the next state $\textupsilon(t + 1)$. Note that this relationship accounts for the amount of energy drained $\theta_{\rm MEC}(t)$, that harvested $H(t)$ and that purchased from the electrical grid $E(t)$, which together lead to the next buffer level $B(t+1)$ through Eq.~\eq{eq:offgrid}. 
The online resource management algorithm, ARCES, finds the best control action vector that yields the desired energy savings within the computing environment. Specifically, for each time slot $t$, problem \eq{eq:objt} is solved, obtaining control actions for the prediction horizon $T$. The control action that is applied at time $t$ is $\psi^{*}(t)$, which is the first one in the retrieved control sequence. This control amounts to setting the number of instantiated \acp{VM}, $M^*(t)$ (along with their obtained $\{\alpha_m^{*}(t)\}$, $\{P_m^{*}(t)\}$, $\{\lambda_m^{*}(t)\}$ values), NIC status to either active or not, $\zeta^{*}(t) \in (0,1)$, and the optimal transmission drivers, $Y^{*}(t)$. The entire process is repeated every time slot $t$ when the controller can adjust the behavior given the new state information. 

Since the actual values for the system input cannot be measured until the next time instant when the controller adjusts the system behavior, the corresponding system state for $t+1$ can only be estimated as:
\begin{equation}
       \hat{\textupsilon}(t + 1) = \phi(\textupsilon(t),\psi(t)) \,.
       \label{eq:state_forecast}
\end{equation}
For these estimations we use the forecast values of load $\hat{L}_{\rm in}(t)$ and harvested energy $\hat{H}(t)$, from the LSTM forecasting module.\\

\subsubsection*{\bf Remark 1 (VM provisioning and load distribution)} a remark on the provisioned VMs at slot $t$, $M(t)$, is in order. The number of active \acp{VM} depends on the forecasted server workload, $\hat{L}_{\rm in}(t+1)$, and each \ac{VM} can compute an amount of up to $\lambda_{\max}$ (considering that virtualization technologies specify the minimum and maximum amount of resources that can be allocated per VM~\cite{migrationpower}). Then, the projected number of VMs that shall be active in slot $t$ to serve the forecasted server workloads is hereby obtained as: \mbox{$M(t) = \ceil[\big] {(\hat{L}_{\rm in}(t+1)/ \lambda_{\max})}$}, where $\ceil[\big] {\cdot}$ returns the nearest upper integer. We heuristically split the workload among VMs by allocating a workload $\lambda_m(t) = \lambda_{\max}$ to the first $M(t)-1$ \acp{VM}, $m=1,\dots,M(t)-1$, and the remaining workload $\lambda_{m}(t) = \hat{L}_{\rm in}(t+1) - (M(t)-1)\lambda_{\max}$ to the last one. This load distribution is motivated by the shares feature~\cite{migrationpower} that is inherent in virtualization technologies. This enables the resource scheduler to efficiently distribute resources amongst contending \acp{VM}, thus guaranteeing the completion of the computation process within the expected time.\\

\subsubsection{The ARCES algorithm}
\label{alg}

\begin{small}
\begin{algorithm}[t]
\begin{tabular}{l l}
{\bf Input:}  & $\textupsilon(t)$ (current state) \\
{\bf Output:} & $\psi^{*}(t)$  (control input vector)\\
01:		& \hspace{-1cm} Parameter initialization\\
		& \hspace{-1cm} ${\mathcal S}(t) = \{\textupsilon(t)\}$ \\
02:		& \hspace{-1cm} {\bf for} ($n$ within the prediction horizon of depth $T$) {\bf do}\\
		& \hspace{-1cm}\quad - $\hat{L}_{\rm in}(t+n)$:= forecast the workload  \\
		&\hspace{-1cm}\quad - $\hat{\rm H}(t+n)$:= forecast the energy\\
		& \hspace{-1cm}\quad - ${\mathcal S}(t+n) = \emptyset$ \\
03:		& \hspace{-1cm}\quad {\bf for} (each $\textupsilon(t)$ in ${\mathcal S}(t+n)$) {\bf do}\\
             & \hspace{-1cm}\qquad - generate all reachable states $\hat{\textupsilon}(t+n)$\\
             & \hspace{-1cm}\qquad - ${\mathcal S}(t+n) = {\mathcal S}(t+n) \cup \{\hat{\textupsilon}(t+n)\}$\\
04:		& \hspace{-1.1cm} \quad\quad {\bf for} (each $\hat{\textupsilon}(t+n)$ in $\mathcal S(t+n)$) {\bf do}\\
            & \hspace{-1.1cm}\qquad\quad - calculate the corresponding $\theta_{\rm MEC}(\hat{\textupsilon}(t+n))$\\
		& \hspace{-1.1cm} \quad\quad {\bf end for}\\
		& \hspace{-1.1cm}\quad\quad {\bf end for}\\
		& \hspace{-1cm} \quad {\bf end for}\\
05:		& \hspace{-1cm} - obtain a sequence of reachable states yielding\\
        & \hspace{-1cm}\quad minimum energy cost\\		
06:		& \hspace{-1cm} {$\psi^{*}(t):=$ control leading from $\textupsilon(t)$ to $\hat{\textupsilon}_{\min}$}\\
07:		& \hspace{-1cm} {\bf Return $\psi^{*}(t)$}
\end{tabular}
\caption{ARCES Pseudocode}
\label{algo:arces}
\end{algorithm}
\end{small}

in order to obtain the best control action that will adjust the computing system behavior at time $t$, with negligible computational overhead, the controller explores the prediction horizon of comprising discrete states and comes up with the feasibility action set that yields the minimum energy cost, i.e., $\psi(t) = (M(t),\{\alpha_m(t)\},\{P_m(t)\},\{\lambda_m(t)\},\zeta(t),Y(t))$.

The algorithm pseudocode is outlined in Algorithm~\ref{algo:arces} and it follows the technique from~\cite{hayes}. Starting from the {\it initial state}, the controller constructs, in a \mbox{breadth-first} fashion, a tree comprising all possible future states up to the prediction depth $T$. 
The algorithm proceeds as follows: A search set $\mathcal S$ consisting of the current system state is initialized (line 01), and it is accumulated as the algorithm traverse through the tree (line 03), accounting for predictions, accumulated workloads at the output buffer, past outputs and controls. The set of states reached at every prediction depth $t+n$ is referred to as $\mathcal S(t+n)$ (line 02). 
Given $\textupsilon(t)$, we first estimate the workload $\hat{L}_{\rm in}(t+n)$ and harvested energy $\hat{H}(t+n)$ (line 02), and generate the next set of reachable control actions by applying the input workload and energy harvested (line 03). 
The energy cost function corresponding to each generated state $\hat{\textupsilon}(t+n)$ is then computed (line 04). Once the prediction horizon  is explored, a sequence of reachable states yielding minimum energy consumption is obtained (line 05). The control action $\psi^{*}(t)$ corresponding to $\hat{\textupsilon}(t+n)$ (the first state in this sequence) is provided as input to the system while the rest are discarded (line 06). The process is repeated at the beginning of each time slot $t$. 

\section{Performance Evaluation}
\label{sec:per_results}

This section present some selected numerical results of the ARCES algorithm for real server workloads. 
The parameters that were used for the simulations are listed in Table~\ref{tab_opt}. 

\begin{table}
	\caption{System Parameters.}
	\center
	\begin{tabular} {|l| l|l|}
		\hline 
		{\bf Parameter} & {\bf Value} \\ 
		\hline
		Max. number of \acp{VM}, $M$ &  $10$\\
		Min. number of \acp{VM}, $d$ & $1$ \\
		Time slot duration, $\tau$   &  $\SI{1} {\minute}$ \\
		Idle state energy for VM $m$, $\theta_{{\rm idle}, m}(t)$ & $\SI{10} {\joule}$\\
		Max. energy for VM $m$, $\theta_{{\max}, m}(t)$ & $\SI{60} {\joule}$\\
		\mbox{per-VM} reconfiguration cost, $\kappa_e$ & $\SI{0.005} {\joule}/(\SI{} {\mega\hertz})^{2}$\\
		TOE in idle state, $\theta_{\rm idle}^{\rm TOE}(t)$ & $13.1\SI{} {\joule}$\\ 
		Max. allowed processing time, $\Delta$ & $\SI{0.8} {\second}$\\
		Processing rate set, $\{f_m(t)\}$  & $\{0,50,70,90,105\}$\\
		Bandwidth, $W_m$ & $\SI{1} {\mega\hertz}$\\
		Max. number of drivers, $Y$ & $6$\\
		Max. tolerable delay, $\tau_{\rm max}$ & $\SI{2} {\second}$\\
		Noise spectral density, $N_0^{(m)}$ & -$174$ dBm/$\SI{}{\hertz}$\\
		Max. VM $m$ load, $\gamma^{\max}$ & $5$ Mbit\\
		Driver energy, $O_{{\rm opt},y}(t)$ & $\SI{1} {\joule/\second}$\\
		Target transmission rate, $r_{0}$ & $1$ Mbps\\
		Controllable factor of delay, $\alpha$ & $0.96$\\
		Reconfiguration overhead, $\sigma$ & $\SI{20} {\milli\second}$\\
		Energy storage capacity, $\beta_{\rm max}$ & $\SI{490} {\kilo\joule}$\\
		Lower energy threshold, $\beta_{\rm low}$  & $30$\% of $\beta_{\rm max}$\\
		Upper energy threshold, $\beta_{\rm up}$  & $70$\% of $\beta_{\rm max}$\\
		\hline 
	\end{tabular}
	\label{tab_opt}
\end{table}

\subsection{Simulation Setup}

\begin{figure}[t]
	\centering
	\resizebox{\columnwidth}{!}{\input{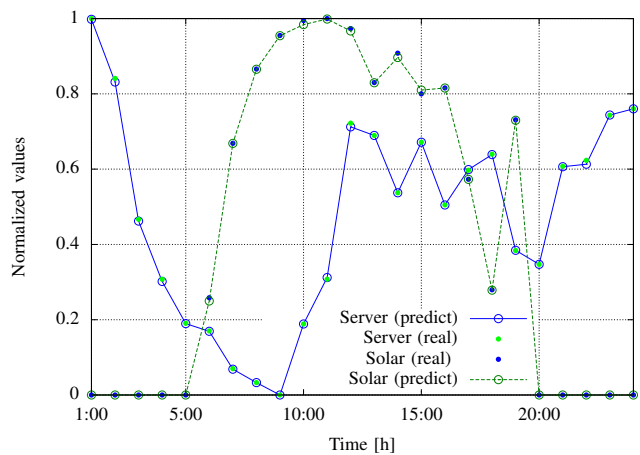}}	
	\caption{Forecast mean value for $L(t)$ and $H(t)$.}
	\label{fig:bs_load}
\end{figure} 

As one of the MEC deployment scenarios, we assume that the MEC server is placed at an aggregation point where \acp{BS} in proximity can offload their computation workload following the random \mbox{real-valued} arrival process. 
%
Our time slot duration $\tau$ is set to $\SI{1} {\minute}$ and the time horizon is set to $T = 3$ time slots. The simulations are carried out by exploiting the Python programming language.

\subsection{Numerical Results}
\label{sub:results}

In Fig.~\ref{fig:bs_load}, we show real and predicted values for the server workloads (Server) and harvested energy (Solar) over time. We track the \mbox{one-step} predictive mean value at each step of the online forecasting routine. The obtained average prediction error (RMSE) for the server workloads and harvested energy processes, both normalized in [0,1] for $T \in \{1,2,3\}$, are $L_{\rm in}(t) = \{0.017, 0.019, 0.021\}$ and $H(t) = \{0.038, 0.039,0.039\}$. Note that the predictions for $L_{\rm in}(t)$ are more accurate than those of $H(t)$ (confirmed by comparing the average RMSE), due to differences in the used dataset granularity. However, the measured accuracy is deemed good enough for the proposed optimization.

\begin{figure}
	\centering
	\resizebox{\columnwidth}{!}{\input{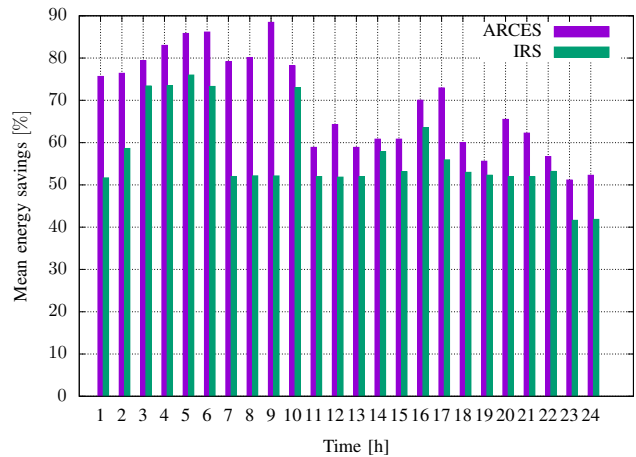}}
	\caption{Mean energy savings within the MEC server.}
	\label{fig:energy_sav}
\end{figure} 
\begin{figure}
	\centering
	\resizebox{\columnwidth}{!}{\input{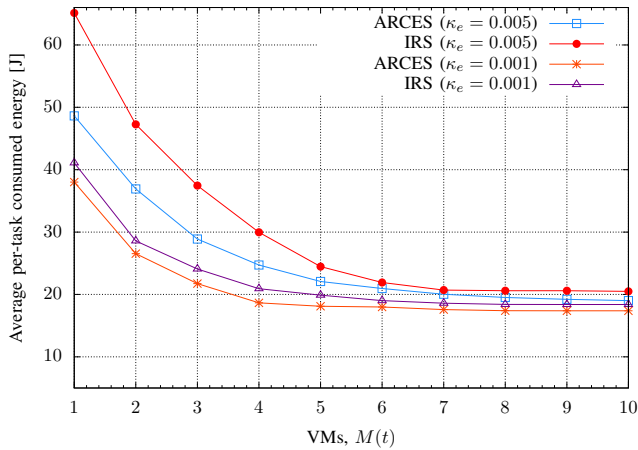}}
	\caption{Average per-task consumed energy vs VMs.}
	\label{fig:vm_cons}
\end{figure}

Our online server management algorithm (ARCES) is benchmarked with another one, named \mbox{Iterative-based} Resource Scheduler (IRS), which is inspired by the iterative approach from~\cite{shojafar2015energy}. In IRS, the optimum \mbox{computing-plus-communication} parameters are obtained in an iterative manner: at the end of each cycle, convergence conditions are used to determine if the found solution is acceptable or the optimization process should continue. The observed conditions are as follows: (i) to ensure that the total load $L_{\rm in}(t)$ has been fully allocated with accuracy $\frac{\sum_{m = 1}^{M(t)} \lambda_m(t)- L_{\rm in}(t)}{L_{\rm in}(t)} \leq \epsilon$, with $\epsilon = 0.01$; (ii) to verify if the selected working frequency $f_m(t)$ is able to cope with the input load $L_{\rm in}(t)$, guaranteeing that the computation processing time is within the server's response time limit $\Delta $, i.e., $L_{\rm in}(t) \leq f_m(t)\Delta$ .
The average energy savings obtained by ARCES are shown in Fig.~\ref{fig:energy_sav}. The average results for ARCES ($\kappa_e = 0.005, \Gamma_m = \SI{0.5} {\milli\watt}, \eta = 1.4$ Gbit/$\SI{} {\joule}$) show energy savings of $69$\%, while IRS achieves $56$\% on average, in both cases with respect to the case where no energy management procedures are applied; i.e., the MEC server provisions the computing resources for maximum expected computation workload (maximum value of $\theta_{\rm MEC}(t)$, with $M=10$, $\forall t$). As expected, the highest energy savings peak is observed at $\SI{9} {\hour}$ as the aggregate computation requests/workload was at its lowest with an expected increase in the computation workload and harvested solar energy in the near future. The effectiveness of the joint VM \mbox{soft-scaling} and tuning of transmission drivers, coupled with foresighted optimization is observed in the obtained numerical results.

In Fig.~\ref{fig:vm_cons}, we show the effects of the \mbox{per-VM} reconfiguration cost on $\theta_{\rm MEC}(t)$ at $\kappa_e = \{0.001,0.005\}$ ($\Gamma_m = \SI{0.5} {\milli\watt}, \eta = 1.4$ Gbit/$\SI{} {\joule}$), taking into account the performance of ARCES when compared with IRS for $M(t) = 1, \dots, M$.
It can be observed that $\theta_{\rm MEC}(t)$ increases with large $\kappa_e$ only for small values of $M(t)$, and as $M(t)$ increases the energy consumption decreases for large $\kappa_e$. Moreover, ARCES leads to an energy consumption reduction with respect IRS from $25$\% to $7$\% (case of $\kappa_e = 0.005$) and from $7$\% to $5$\% (case of $\kappa_e = 0.001$). When ARCES is compared with the case where no energy management is applied (maximum value of $\theta_{\rm MEC}(t)$, with $M=10$, $\forall t$), the obtained energy reduction ranges from $45$\% to $31$\% (case of $\kappa_e = 0.005$) and from $25$\% to $21$\% (case of $\kappa_e = 0.001$).
These numerical results confirm that jointly autoscaling the available \mbox{computing-plus-communication} resources within the computing platform provides remarkable energy savings. These results conforms to our expectations from~\cite{nicola}.  

\section{Conclusions}
\label{sec:concl}

In this paper, we have envisioned a \mbox{hybrid-powered} MEC server placed in proximity to a \ac{BS} cluster for handling the offloaded computation workload. Moreover, the use of green energy promotes energy \mbox{self-sustainability} within the network. 
We have considered a \mbox{computing-plus-communication} energy model, within the {\it\ac{MEC}} paradigm, and then have put forward a combination of a traffic engineering- and \ac{MEC} Location \mbox{Service-based} online server management algorithm with {\it \ac{EH}} capabilities, called Automated Resource Controller for \mbox{Energy-aware} Server (ARCES), for autoscaling and reconfiguring the \mbox{computing-plus-communication} resources. The main goal is to minimize the overall energy consumption, under hard \mbox{per-task} delay constraints (i.e., \ac{QoS}).  
ARCES jointly performs (i) a \mbox{short-term} server demand and harvested solar energy forecasting, (ii) \ac{VM} \mbox{soft-scaling}, workload and processing rate allocation and lastly, (iii) switching on/off of transmission drivers (i.e., {\it fast} tunable lasers) coupled with the \mbox{location-aware} traffic scheduling.
Numerical results, obtained with \mbox{real-world} energy and server workload traces, demonstrate that the proposed algorithm (ARCES) achieves energy savings of $69$\%, on average, with an energy consumption ranging from $31$\%-$45$\% at high \mbox{per-VM} reconfiguration cost and from $21$\%-$25$\% at low \mbox{per-VM} reconfiguration cost, with respect to the case where no energy management techniques are applied. 

\section*{Acknowledgements}

This work has received funding from the European Union's Horizon 2020 research and innovation programme under the Marie Sklodowska-Curie grant agreement No. 675891 \mbox{(SCAVENGE)}.

\bibliographystyle{IEEEtran}
\scriptsize
\bibliography{biblio}
\end{document}